\documentclass[aps,prl,amssymb,showpacs]{revtex4}
\begin{document}
\title{Ten themes of viscous liquid dynamics}
\author{Jeppe C. Dyre,}
\address{DNRF Centre ``Glass and Time,'' IMFUFA (Building 27), Department of Science, 
Roskilde University, Postbox 260, DK-4000 Roskilde, Denmark}
\date{\today}

\begin{abstract}
Ten ``themes'' of viscous liquid physics are discussed with a focus on how they point to a general description of equilibrium viscous liquid dynamics (i.e., fluctuations) at a given temperature. This description is based on standard time-dependent Ginzburg-Landau equations for the density fields, stress tensor fields, potential energy density field, and fields quantifying molecular orientations. One characteristic aspect of the theory is that density has the appearance of a non-conserved field. Another characteristic feature is long-wavelength dominance of the dynamics, which not only simplifies the theory by allowing for an ultra-local Hamiltonian (free energy), but also explains the observed general independence of chemistry. Whereas there are no long-ranged static (i.e., equal-time) correlations in the model, there are important long-ranged dynamic correlations on the alpha time scale.
\end{abstract}


\maketitle

\section{Introduction}

Glass-forming liquids approaching the calorimetric glass transition become extremely viscous, with viscosity $10^{13}-10^{15}$ times than that of ambient water. These extreme viscosities reflect the fact that molecular motion has almost come to a standstill. Most molecular motion is vibrational in these liquids, thus a computer simulation of a liquid close to the calorimetric glass transition with present-day computers would not be able to distinguish it from a disordered solid. Long time ago Kauzmann and Goldstein recognized that most motion is vibrational \cite{kau48,gol69}, in fact already around 1930 the important German glass scientist Simon seems to have been aware of this \cite{sim31}. Molecular reorientations do take place at rare occasions by a sudden and (presumably) fairly localized jump from one solid structure to another resulting in a reordering of the molecules. Kauzmann described these rare {\it flow events} as ``jumps of molecular units of flow between different positions of equilibrium in the liquid's quasicrystalline lattice'' \cite{kau48}, thus emphasizing the fact that the liquid is much like a solid. In a series of papers \cite{dyr99,dyr05a,dyr05b,dyr06a} we suggested the word ``solidity'' for this property and conjectured that viscous liquids are qualitatively different from less viscous liquids and more to be thought of as ``solids that flow''. 

In the present paper we summarize several postulates, claims, conjectures -- more neutrally: ``themes'' -- relating to viscous liquid dynamics in general and to solidity in particular. Some of these themes are fairly uncontroversial, but several of them challenge the prevailing views in the research field. After presenting the themes we summarize a proposal for describing viscous liquid dynamics at a given temperature. Finally, a brief discussion is given.

\section{General themes}

This section summarizes some points which, although perhaps not generally agreed upon, should not be very controversial for the general glass community.

{\it Theme 1: The three non's.}
Much of the attraction of the study of viscous liquids and the glass transition lies in the universal physical properties of highly viscous liquids \cite{bra85,sch86,ang00,deb01,dyr06b}. Glass formation is a universal property: It is generally agreed that (with liquid Helium as the outstanding exception) all liquids form glasses if cooled rapidly enough to avoid crystallization. From purely macroscopic measurements it is not possible to distinguish chemically quite different liquids (covalently bonded, ionically bonded, bulk metallic glass formers, hydrogen bonded, van der Waals bonded). With only few exceptions each of these classes of liquids generally show {\it non-exponential} relaxation functions (linear as well as nonlinear), {\it non-Arrhenius} temperature dependence of the average relaxation time with an activation energy that increases upon cooling, and {\it nonlinearity} of responses after even small temperature jumps (e.g., 1\%). 

{\it Theme 2: Separating the non-exponential problem from the non-Arrhenius problem.}
Most theories for viscous liquid dynamics and the glass transition attempt to explain both the non-exponentiality and the non-Arrhenius temperature dependence in one single theory. This is motivated by the generally accepted view that fragility correlates with the $\beta$ exponent of the stretched exponential \cite{boh93}, thus indicating a deep connection between the two {\it non's}. In our view -- to some extent based on measurements performed in our laboratory during the last 20 years on simple, organic glass-forming liquids -- there is no such clear correlation. Whether or not one accepts this, it seems to be a reasonable strategy to approach a difficult problem by ``slicing the Gordian Knot,'' and this is what we propose to do. Thus we here ignore the non-Arrhenius problem (see, e.g., the recent review \cite{dyr06b}) and focus exclusively on describing the viscous liquid's equilibrium fluctuations at a given temperature: A description is sought where the dynamic parameters are unspecified functions of temperature. 

{\it Theme 3: Inherent dynamics.}
As mention, there is consensus that the dynamics of highly viscous liquids consist of rare jumps over large potential energy barriers, ``flow events''.  Moreover, the consensus is that it is a good approximation to assume that the vibrational dynamics decouple from, and is independent of, the configurational dynamics resulting from flow events. The standard picture is that the configurational dynamics freeze in the glass (with the exception of beta processes that, however, are unable to induce a macroscopic flow or relax a macroscopic stress), whereas the vibrational dynamics continue in the glass. This picture of the glass transition explains the fact that specific heat and thermal expansion coefficient are smaller in the glass than in the liquid, because the configurational degrees of freedom only contribute to these quantities in the liquid phase. This old explanation is undoubtedly basically correct. (There is one problem, though: It is usually assumed that the phonon spectrum has a similar temperature dependence in the liquid and in the glass, which is incorrect. This has misled people to think that the configurational contribution to the specific heat may be estimated by subtracting the extrapolated glass specific heat into the liquid phase. This is wrong because the high-frequency elastic constants are usually much more temperature dependent in the liquid than in the glass \cite{dyr06b}. Consequently, the vibrational entropy is much more temperature dependent in the liquid phase than in the glass.)

The above picture was confirmed during the last ten years by numerous computer simulations. The picture that emerges is one where the potential energy minima in the high-dimensional energy landscape play a crucial role. This was the vision of Goldstein already in 1969 \cite{gol69}, which was further elaborated in works by Stillinger and Weber \cite{sti83}. The latter authors introduced the concept of an ``inherent structure,'' the basin of attraction in configuration space of a given potential energy minimum. The low-temperature dynamics are thought of as vibrations around the potential energy minima, interrupted by occasional jumps between basins. The latter jumps are obviously the flow events, the dynamics of which have been referred to as ``inherent dynamics'' \cite{sch00}.

{\it Theme 4: A small number.}
Viscous liquids approaching the calorimetric glass transition are characterized by an extremely small number. To see this, note first that any liquid has several diffusion constants, for instance the heat diffusion constant, the single-particle diffusion constant determining the long-time mean-square displacement, the so-called coherent diffusion constant determining the long-time decay of long-wavelength density fluctuations, and the kinematic viscosity of the Navier-Stokes equation (viscosity over density, the transverse momentum diffusion constant). In ``ordinary'' liquids with viscosity similar to that of ambient water these diffusion constants are all within one or two orders of magnitude from $10^{-7}\rm m^2/s$. This is easy to understand from elementary kinetic theory according to which a diffusion constant is the mean-free length of the diffusing entity squared over the mean time between collisions: Taking as typical microscopic parameters for these quantities one Angstrom and $0.1$ picosecond, respectively, one gets the number $10^{-7}\rm m^2/s$. This estimate, however, completely breaks down for highly viscous liquids: The heat diffusion constant is relatively unaffected by viscosity. The single-particle diffusion constant becomes extremely small. The kinematic viscosity follows viscosity and becomes extremely large. Thus the ratio between single-particle diffusion constant and kinematic viscosity goes from roughly $1$ to roughly $10^{-30}$. Such an extremely small number is most unusual in condensed matter physics. Usually in physics small numbers imply a simplification of some kind. Thus there is hope that the correct theory of viscous liquid dynamics is fairly simple. Because the high viscosity directly reflects the high energy barriers for flow events, this also seems to have been Goldstein's view when he in his 1969 paper wrote: ``I am only conjecturing that whatever rigorous theory of kinetics we will someday have, processes limited by a high potential barrier will share some common simplifications of approach'' \cite{gol69}.

\section{Themes pointing towards a theory for the dynamics}

The above themes presumably are relatively uncontroversial. We proceed to discuss some more speculative themes that taken together point towards a specific theory of viscous liquid dynamics.

{\it Theme 5: Polymers are different.}
Polymers have a glass transition below which the structure is frozen, just as is the case for the liquid-glass transition. Maybe half of all published data for the glass transition and the dynamics just above $T_g$ are for polymers. There is one crucial aspect, however, where the polymer glass transition differs from the liquid-glass transition, namely in the fact that polymers do not flow above the transition. Here one is in the rubbery regime which is characterized by very long relaxation times due to entanglement effects. In contrast, glass-forming liquids do not have relaxation times much larger than the alpha relaxation time. In fact, one of their very characteristics is a notable sharp long-time cut-off in the relaxation time distribution. This cut-off is seen experimentally in the observation that the low-frequency side of linear relaxation functions (e.g., the dielectric relaxation, shear and bulk modulus measurements) is always virtually Debye. This is not the case for polymers. In our opinion this difference is important, and the polymer glass transition should be treated as a separate issue. In this connection it is interesting to note that plastic crystals, although they also do not flow above $T_g$, seem to behave much more like glass-forming liquids \cite{lun00}.

{\it Theme 6: Back to 3 dimensions.}
The energy landscape approach is a popular way of treating viscous liquids (see, e.g., the recent review by Sciortino \cite{sci05}). The idea is to focus on the motion in configuration space, justified by the fact that the potential energy function completely determines the dynamics. As mention, at high viscosity the slow (alpha) dynamics may be identified with the inherent dynamics taking the system from one energy minimum to another \cite{sch00}. The barriers impeding this motion are saddle points in configuration space. This description is general and undoubtedly correct, but the question is how useful it is. The problem is that the energy landscape is exceedingly complex. In condensed matter physics most phenomena play out differently in two, three and four dimensions. We don't know whether this is also the case for viscous liquid dynamics, but it seems to be a good guess. If this is so, the obvious question is how dimensionality is reflected in the energy landscape. This question is not easy to answer, although it seems not to have been discussed in the literature. Most landscape papers are general and their reasoning could equally well be applied to the protein problem. But if dimensionality does matter, these treatments may miss important characteristics of the problem. The simplest cure seems to be to go back to three dimensions in the modelling.

{\it Theme 7: Solidity.}
The basic idea of inherent dynamics is that the slow dynamics of viscous liquids basically consist of jumps between potential energy minima in configuration space. A potential energy minimum defines a state of mechanical equilibrium, i.e., a state where the force on each molecule is zero. Such a state may be thought of as a disordered solid. Thus we propose to regard a viscous liquid as it develops in time as a sequence of disordered solids with flow events giving instantaneous transitions between two (very similar) solid states. Due to the fact that the sound velocity is finite, this point of view strictly speaking is only realistic below the solidity length $l$ which, if $c$ is the (high-frequency) sound velocity, $\tau$ the average relaxation time, and $a$ the intermolecular distance, is given \cite{dyr99} by

\begin{equation}
l^4\,=\,
c\,\tau\, a^3\,.
\end{equation}
Just above the calorimetric glass transition the solidity length $l$ is close to $10,000 {\rm\AA}$. 

In the remainder of this paper we shall limit the discussion to the dynamics below the solidity length. The basic idea of regarding a viscous liquid as a time-sequence of disordered solids may be summarized as follows:

\begin{equation}\label{1}
{\rm Viscous\,\, liquid\,\,\cong\,\,Solid\,\, that\,\, flows}\,.
\end{equation}
The conjecture is that viscous liquids are qualitatively different from the less-viscous liquids studied in conventional liquid state theory \cite{boonyip,hanmc}. 

{\it Theme 8: Solidity implies apparent density and momentum non-conservation.}
Flow events may be regarded as instantaneous on length scales below the solidity length. We now argue that, despite the fact that molecules obviously cannot just appear or disappear, in a correct description of the inherent dynamics density must be treated as a non-conserved field variable. A given flow event takes the system from one potential-energy minimum to another. Far from the flow event centre the displacements are small, obviously, but nevertheless they cannot be ignored. By solving the standard equations of solid elasticity one finds that the leading term of the displacement field far away is a radially symmetric term that varies with distance from the flow event $r$ as $r^{-2}$ \cite{dyr06a}. This is a zero-divergence displacement field and thus there are no density changes far from the flow event. That may sound counterintuitive, but what happens is that for any given spherical shell surrounding the flow event the same number of molecules passes this shell and any other shell (inwards or outwards). A flow event may be regarded as the analogue of Hilbert's hotel, the infinite hotel that even when totally occupied can always house one extra guest (by asking each guest to move to one higher room number). Numbering the flow events consecutively after the time they take place, $t_\mu$, if ${\bf r}_\mu$ is the centre of the $\mu$'th flow event and the number $b_\mu$ measures its magnitude, the above considerations translate into the following dynamic equation for the density $\rho({\bf r},t)$ in a coarse-grained description \cite{dyr06a}:

\begin{equation}\label{2}
\dot\rho({\bf r},t)\,=\,\sum_\mu b_\mu\delta({\bf r}-{\bf r}_\mu)\delta(t-t_\mu)\,.
\end{equation}
This equation does not account for correlations between different flow events, so it does not constitute a theory. Equation (\ref{2}) is just a {\it description} of equilibrium density fluctuations in viscous liquids, a description that reflects the solidity-based fact that density has the appearance of a non-conserved field. Moreover, the description is too rough for a useful theory which -- as we shall see -- must include the tiny density changes in the far field of a flow event (as well as, of course, correlations between different flow events).

Momentum conservation is equally irrelevant in viscous liquids because the transverse momentum diffusion constant, the so-called kinematic viscosity of Navier-Stokes equation, is enormous. Just above the calorimetric glass transition this quantity is approximately a factor $10^{15}$ larger than, e.g., in ambient water. Even at the somewhat higher temperature where the alpha relaxation time is one second, transverse momentum diffuses more than one kilometre on the alpha relaxation time scale. Thus momentum is continuously exchanged with the sample holder, and a description based on momentum conservation {\it may} be misleading. For instance, Newton's second law implies that the centre of mass does not move during a flow event. It was recently proposed that, in fact, the total displacement of all molecular positions $\Delta\bf R$ is an important characteristic of a flow event \cite{dyr06a,note}. 

{\it Theme 9: k-vectors are important because long-ranged elastic effects cannot be ignored.}
A mechanical disturbance of a solid at one point gives rise to elastic deformations spreading to infinity; this is how the solidity of viscous liquids implies that density acquires the appearance of a non-conserved field. Thus the rough description of Eq. (\ref{2}) can come about only in a theory that incorporates long-ranged elastic interactions. The common approach of regarding the liquid as a collection of non-interacting regions 
\cite{ada65,dyr95,sch04,lub06} misses this point and does not properly reflect solidity. One theory which does include long-ranged interactions is the frustration-based approach of Tarjus and co-workers \cite{tar05}, although the frustration-induced interactions are not necessarily of an elastic nature. The simplest way to take long-ranged elastic effects into account is to ensure that these are properly reflected in the {\it k}-vector dependence of the dynamics.

{\it Theme 10: Long-wavelength dominance of the dynamics.}
We now proceed to answer the question: What are the simplest dynamics consistent with the above themes? We choose a field-theoretic description -- field theory is used in virtually every part of modern physics so that seems to be an obvious starting point. Which fields should be included? This question is answered below, but the one field that appears in every field-theoretic description of liquids is the density field; for the moment we limit the reasoning to this field. Which dynamics should it obey? The simplest choice is a standard time-dependent Ginzburg-Landau equation (giving a Langevin equation for each degree of freedom). If $\beta$ is the inverse temperature, $H$ the ``Hamiltonian'' (free energy), and $\xi_{\bf k}^*(t)=\xi_{-\bf k}(t)$ a Gaussian white noise term, in $k$ space the equation looks as follows:

\begin{equation}\label{3}
\dot\rho_{\bf k}\,=\,
-\Gamma_k \frac{\partial(\beta H)}{\partial \rho_{-\bf k}}\,+\,\xi_{\bf k}(t)\,.
\end{equation}
We have not assumed that the rate constant $\Gamma_k$ is independent of $k$. This is because, if the rate constant is not {\it k}-dependent, there is little chance to reproduce the observed relatively broad relaxation time distributions in a simple theory with few free energy minima.

How does $\Gamma_k$ look? From a Taylor expansion one would expect that at small {\it k}

\begin{equation}\label{4}
\Gamma_k\,=\,\Gamma_0+Dk^2\,.
\end{equation}
We shall assume that Eq. (\ref{4}) applies not only for small {\it k} vectors, but for all {\it k}. If density behaved as a conserved field, by definition one would have $\Gamma_0=0$ \cite{hoh77}. Equation (\ref{3}) gives the effective description of the flow-event induced density changes. It is understood that we consider a volume with a fixed number of particles and that the $k$ vectors are those consistent with periodic boundary conditions. The fact that there are only a finite number of molecules translates into a cut-off in {\it k}-space with maximum {\it k} vector, $k_c$, given by $k_c\sim 1/a$ where $a$ is the average intermolecular distance. In order for Eq. (\ref{4}) to give relaxation rates covering several decades for the allowed {\it k} vectors, the following inequality must be obeyed:

\begin{equation}\label{5}
D\,\gg\, \Gamma_0\, a^2\,.
\end{equation}
This inequality expresses an assumption of {\it long-wavelength dominance} of the dynamics in the sense that density fluctuations on length scales much larger than $a$ decay on the alpha time scale. This provides a simple way to understand the rough universality of viscous liquid dynamics (i.e., their surprisingly weak dependence of chemistry). Another consequence of long-wavelength dominance of the dynamics is that the Hamiltonian may be taken to be ultra local, i.e., terms in the Hamiltonian  reflecting spatial density correlations may be ignored. This gives a considerable simplification, which is justified by the fact that observed density correlations are short ranged and thus presumably unimportant when the system is coarse-grained on a length scale longer than $a$.

\section{General theory of viscous liquid dynamics: A proposal}

The above themes inspire an approach to viscous liquid dynamics \cite{dyr05a,dyr05b,dyr06b} that may be summarized in four points: 

\begin{enumerate}
\item The relevant degrees of freedom are fields $\phi^{(1)}({\bf r})$, ...,  $\phi^{(n)}({\bf r})$ defined as: a) the densities of the liquid's different molecules, b) for each molecule a field quantifying the density of the configurational variable reflecting the molecular symmetry, c) the 5 components of the traceless stress tensor (the pressure variable is redundant because pressure fluctuations are not independent of density fluctuations), d) the potential energy density;
\item The Hamiltonian $H$ -- the free energy -- is ultra local and a sum of invariant (scalar) terms up to some even order; 
\item For each field the dynamics are described by a time-dependent Ginzburg-Landau equation:

\begin{equation}\label{6}
\dot \phi^{(j)}_{\bf k}=-\Gamma^{(j)}_k\,\frac{\partial (\beta H)}{\partial \phi^{(j)}_{-{\bf k}}}+\xi^{(j)}_{\bf k}(t)\, ,
\end{equation}
 where $\xi^{(j)}_{\bf k}(t)$ is a standard Gaussian white noise term;
\item For each density field the coefficients of Eq. (\ref{6}) are given by expressions of the form $\Gamma^{(j)}_ k=\Gamma^{(j)}_0+D^{(j)} k^2$ where $D^{(j)}\gg\Gamma^{(j)}_0 a^2$, for all remaining fields the coefficients are $k$-independent: $\Gamma^{(j)}_k=\Gamma^{(j)}_0$.
\end{enumerate}
A consequence of the ultra-locality assumption is that the dynamics at any given temperature are determined from the thermodynamics: The thermodynamics split into vibrational and configurational (inherent) contributions \cite{sci05}, and the inherent part of the free energy defines the Hamiltonian. Thus, except for the unknown temperature dependence and relative weight of the coefficients of the Langevin equations, the above approach implies that equilibrium fluctuations are completely determined by the inherent thermodynamics. Note that the ultra-locality assumption is equivalent to assuming that the static structure factor is independent of the {\it k} vector and equal to its $k\rightarrow 0$ limit. Clearly, this theory appears as crude compared to many others, but this follows from assuming that the dynamics on the alpha time scale are dominated by density fluctuations occurring over much longer length scales than the intermolecular distance $a$.

The rough description Eq. (\ref{2}) is inaccurate because it does not incorporate the density changes in the surroundings of a flow event, and because it is not a theory, but just a {\it description} of the fluctuations. Different flow events cannot be uncorrelated, however, because otherwise the density at a given point would increase or decrease without limits. An obvious question is how correlations are taken into account in Eq. (\ref{3}). The answer is that a time-dependent Ginzburg-Landau equation automatically assures consistency with statistical mechanics, and that the gradient term implies that flow events resulting in a free energy increase are less likely (per unit time) than those resulting in a free energy decrease. This induces correlations between flow events and automatically limits density fluctuations at any given point in space to those consistent with the macroscopic compressibility.

\section{How does the theory compare to experiment?}

The above scheme is general. It does not lead to specific experimental predictions because the details of the dynamics depend on the Hamiltonian. For a given Hamiltonian the equations of motion Eq. (\ref{6}) allow calculation of the dynamic structure factor (at wavelengths long enough that the static structure factor may be regarded as constant), as well as of the frequency dependence of the dielectric constant, bulk modulus, shear modulus, and specific heat. 

The simplest non-trivial Hamiltonian has a second-order free-field term and a perturbing third order term. This case has been solved to second order in the perturbation \cite{dyr05b,dyr06a}. Both for the dielectric loss and for the imaginary (loss) part of the bulk modulus the theory predicts losses varying with frequency as $\omega^{-1/2}$ for $\omega\tau\gg 1$ where $\tau=1/\Gamma_0$ is the alpha relaxation time, whereas the losses are predicted to follow the Debye prediction ($\propto\omega$) on the low frequency side of the loss peak ($\omega\tau\ll 1$). This prediction appears to be consistent with experiment (but more work is needed): Most data are fitted by stretched exponentials with exponents $\beta$ in the range 0.3-0.7. Moreover, based on data for ten organic liquids it was proposed in 2001 that whenever time-temperature superposition applies accurately (whenever low-lying Johari-Goldstein beta processes do not interfere with the alpha process), the high-frequency loss is close to $\omega^{-1/2}$ \cite{ols01}.

\section{Outlook}

The themes proposed here as central for understanding viscous liquid dynamics point to a simple framework for the dynamics, a framework based on a standard time-dependent Ginzburg-Landau equation. Ultralocality of the Hamiltonian simplifies things considerably; it follows from the assumption of long-wavelength dominance of the dynamics, an assumption that may be justified by various theoretical arguments \cite{dyr05a,dyr06a}. The latter, of course, would also be consistent with a more accurate Hamiltonian that includes the well-known static correlations over short length scales. -- The proposed framework appears to be simpler than many contemporary theories (as mentioned, most theories attempt to simultaneously solve both the non-Arrhenius and the non-exponential problem). Therefore, it seems worthwhile to pursue this approach until it may be proven inadequate. 

\section{Acknowledgments}
The author wishes to acknowledge useful discussions with Shankar Das, Gregor Diezemann, Giancarlo Ruocco, and from the Roskilde glass group: Nick Bailey, Tage Christensen, Kristine Niss, Niels Boye Olsen, and Thomas Schr{\o}der. This work was supported by the Danish National Research Foundation Centre for Viscous Liquid Dynamics ``Glass and Time.''


\begin{thebibliography}{99}
\bibitem{note} Nick Bailey pointed out that the definition of the total molecular displacement induced by a flow event, $\Delta\bf R$, as the sum of all position changes is not unambiguous. Thus the sum of all position changes depends on the boundary conditions, even in the limit where the boundaries are moved to infinity. A possible solution to this problem is to define $\Delta\bf R$ indirectly via the $k$ component of the density change $\delta\rho_{\bf k}$ as follows (where $\rho_{\bf k}$ is defined in Ref. \cite{dyr06a}, $N$ is the number of molecules, and the averaging is over different positions of the flow event): $N\langle|\delta\rho_{\bf k}|^2\rangle=b^2+(\Delta{\bf R}\cdot{\bf k})^2$ for $k\rightarrow 0$. Independence of the boundary condition is ensured by introducing a convergence factor $\lim_{\Lambda\rightarrow 0}\exp(-\Lambda r)$ where $r$ is the distance to the flow event, and by letting $\Lambda\rightarrow 0$ only after the volume goes to infinity.
\end{thebibliography}
\end{document}